\documentclass[twocolumn,printnumbers,amsmath,amssymb,prl]{revtex4}
\usepackage{graphicx}
\usepackage{color}

\begin{document}

\title{Self-assembling two-dimensional quasicrystals in simple systems of monodisperse soft-core disks}


\author{Mengjie Zu$^1$, Peng Tan$^2$, and Ning Xu$^1$}

\affiliation{$^1$CAS Key Laboratory of Soft Matter Chemistry, Hefei National Laboratory for Physical Sciences at the Microscale, and Department of Physics, University of Science and Technology of China, Hefei 230026, People's Republic of China.\\
$^2$State Key Laboratory of Surface Physics and Department of Physics, Fudan University, Shanghai 200433, People's Republic of China.}

\begin{abstract}
In previous approaches to form quasicrystals, multiple competing length scales involved in particle size, shape or interaction potential are believed to be necessary. It is unexpected that quasicrystals can be self-assembled by monodisperse, isotropic particles interacting via a simple potential without multiple length scales. Here we report the surprising finding of the self-assembly of such quasicrystals in two dimensional systems of soft-core disks interacting via repulsions. We find not only dodecagonal but also octagonal quasicrystals, which have not been found yet in soft quasicrystals. In the self-assembly of such unexpected quasicrystals, particles tend to form pentagons, which are essential elements to form the quasicrystalline order. Our findings pave an unexpected and simple way to form quasicrystals and pose a new challenge for theoretical understanding of quasicrystals.
\end{abstract}

\maketitle

Quasicrystal (QC) is a fantastic discovery in materials science and condensed matter physics \cite{shechtman,levine}, which exhibits a rotational symmetry forbidden in periodic crystals. Since the first observation of a decagonal QC in Al-Mn alloys \cite{shechtman}, thousands of metallic QCs have been obtained \cite{steurer}. These QCs intrinsically involve multiple length scales arising from multi-type atoms. Soft or mesoscopic (non-metallic) QCs have brought great attentions to the community of QCs recently \cite{hayashida,talapin,lee,fischer,xiao,wasio,ye} since the first finding of a $12$-fold QC in supramolecular dendrimers \cite{zeng}. Compared with metallic QCs, soft materials have displayed advantages in forming stable mono-component QCs. However, multiple length scales still seem to be inevitable to form soft QCs. Up to now, soft QCs are obtained by either introducing multiple competing length scales in the inter-particle potential \cite{dzugutov,engel,iacovellaa,archer,dotera,engel1} or using anisotropic particles naturally possessing multiple length scales, such as tetrahedral and patchy particles \cite{haji,reinhardt}. It is unexpected to self-assemble QCs by mono-component, isotropic particles interacting via a smooth potential without involving multiple length scales.

Here we show that such an unexpected self-assembling of soft QCs do exist in high-density systems containing monodisperse, soft-core disks interacting via a simple pairwise repulsion, $U(r)=\frac{\epsilon}{\alpha}\left( 1-r/\sigma\right)^{\alpha}\Theta\left(1-r/\sigma\right)$, where $r$ is the separation between two disks, $\sigma$ is the disk diameter, $\epsilon$ is the characteristic energy scale, $\alpha$ determines the softness of the potential, and $\Theta(x)$ is the Heaviside step function. With increasing number density $\rho$ at fixed temperature $T$, solid phases with different structures emerge in sequence, as shown in Fig.~1a. Figure~1b shows that the inter-particle potential does not exhibit multiple length scales. Surprisingly, in certain ($\rho, \alpha$) parameter regimes, both octagonal and dodecagonal QCs (OQCs and DDQCs) appear. To our knowledge, OQCs have not yet been convincingly observed in soft QCs. To avoid clustering of particles \cite{miyazaki}, we vary $\alpha$ from $2$ to $3$.

\begin{figure*}
\includegraphics[width=0.95\textwidth]{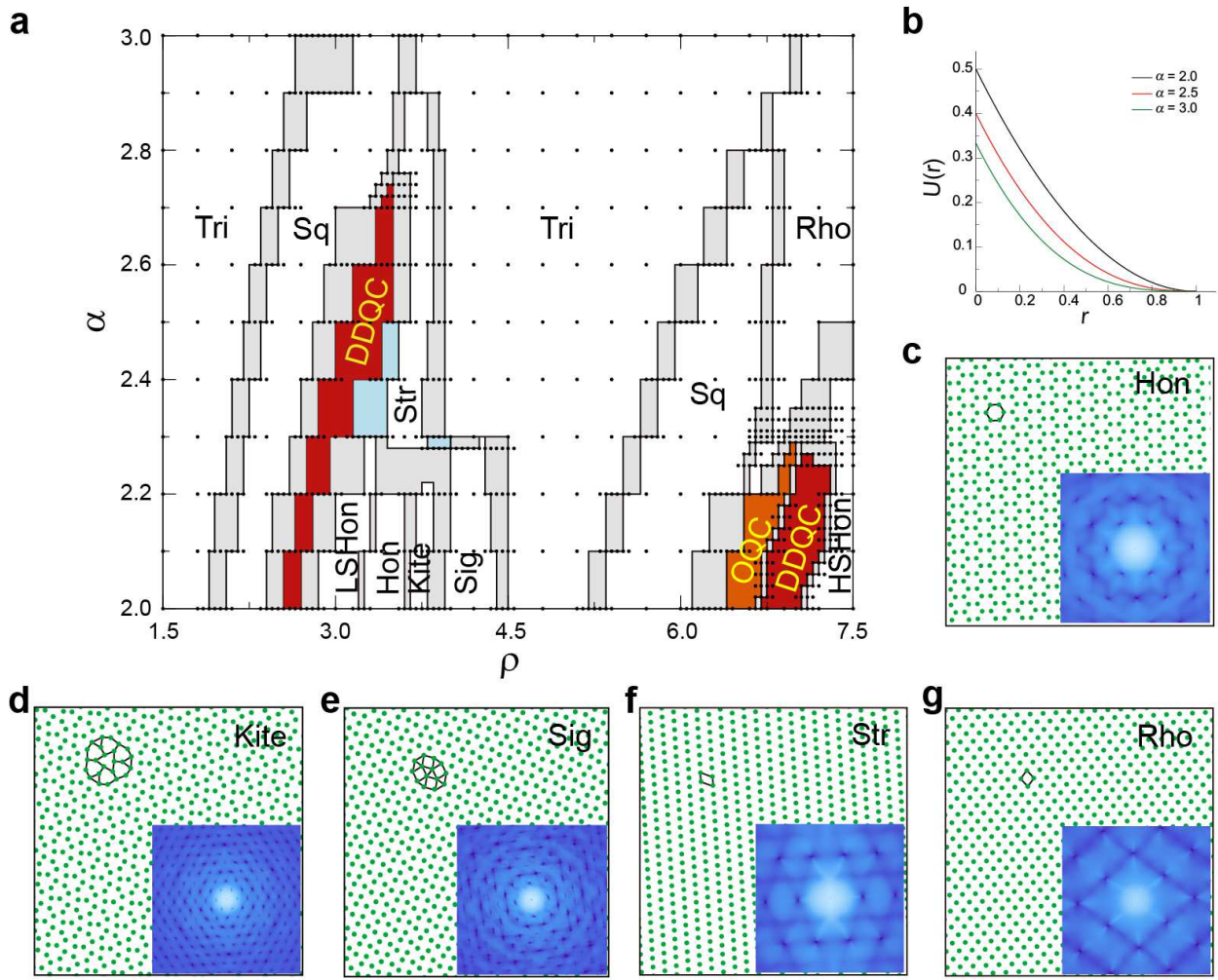}
\caption{\label{fig:fig1} {\bf{Multiple solid phases self-assembled by soft-core disks at high densities.}} {\bf{a}}, Phase diagram of solid states in terms of number density $\rho$ and potential exponent $\alpha$ at a fixed temperature $T=10^{-4}$. The black dots label the $(\rho,\alpha)$ pairs where we run the simulations to identify states. The orange and red areas are the territories of OQCs and DDQCs, respectively. The gray and light blue areas are phase coexistence regimes of two and three types of solids, respectively. In addition to the abbreviations defined in the text, LSHon and HSHon denote lower- and higher-density stretched honeycomb, respectively.  {\bf{b}}, Examples of particle interaction potentials with different $\alpha$. {\bf{c-g}}, A part of static configurations of five distinct crystalline solids with particle diameters shown here being $28\%$ of their actual values. The ($\rho, \alpha$) values of the five states are (3.35,2.0), (3.70,2.0), (3.95,2.0), (3.60,2.3), and (6.90,2.4), respectively. The solid lines outline the unit cell of each solid. The insets are diffraction patterns.
}
\end{figure*}

In Figs.~1c-1g, we first show the static configuration and diffraction pattern of five special crystals other than the ordinary triangular and square solids, including honeycomb (Hon), kite (Kite), sigma-phase (Sig), stripe (Str), and rhombus (Rho) solid. Each solid has a definite unit cell as outlined in the configuration. Although some unit cells are complicated, they repeat periodically in space, leading to a periodic diffraction pattern.

QCs exist in three isolated regimes of the phase diagram. OQCs occupy a regime with small $\alpha$ and high $\rho$. DDQCs emerge in two regimes: One adjacent to OQCs and the other at relatively low $\rho$, covering a wider range of $\alpha$.

\begin{figure*}
\includegraphics[width=0.95\textwidth]{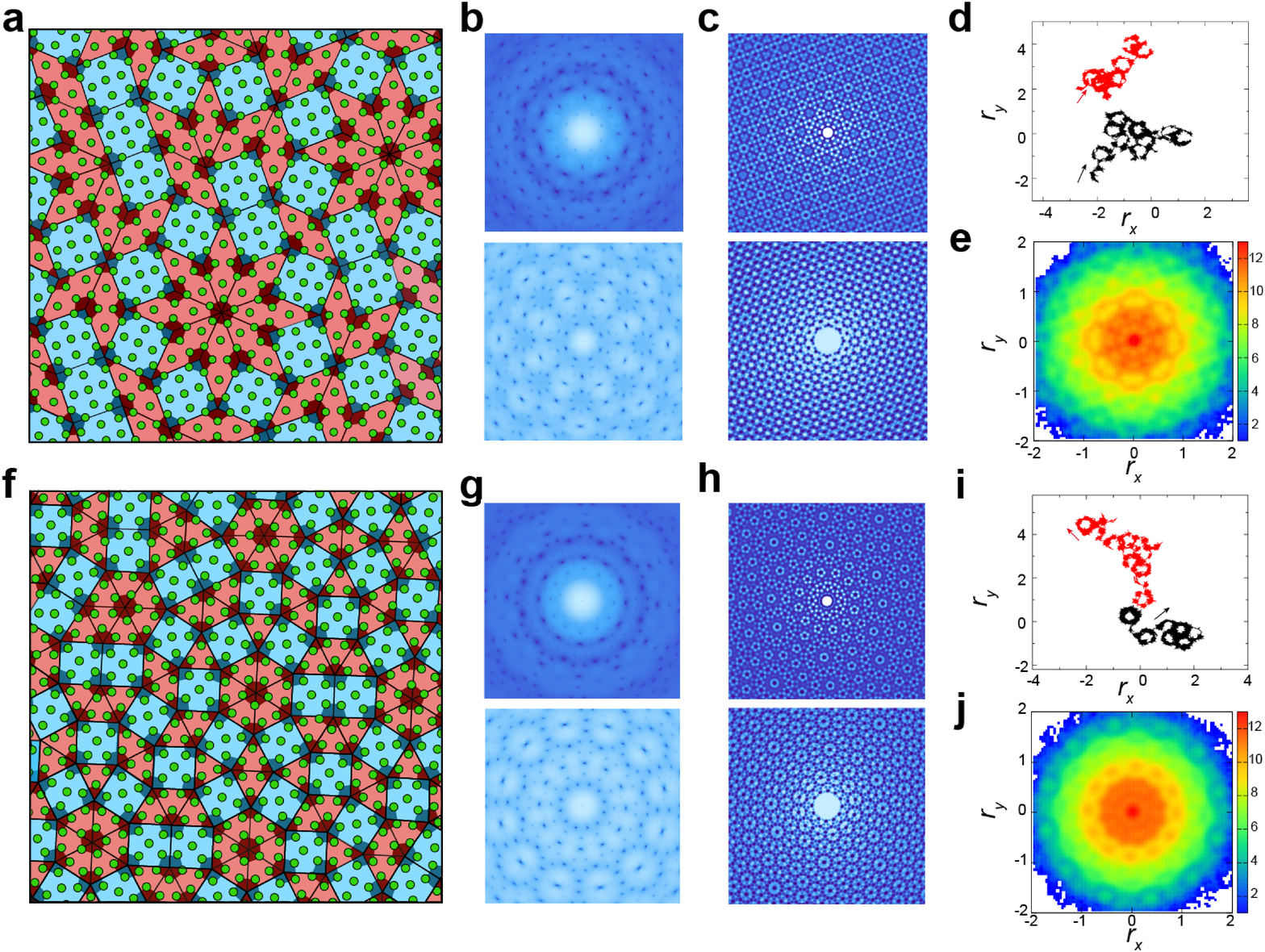}
\caption{\label{fig:fig2}  {\bf{Characterization of structure and dynamics of two types of quasicrystals.}} Top and bottom rows are for a OQC and DDQC at $(\rho, \alpha)=(6.6, 2.0)$ and $(7.0, 2.0)$, respectively. {\bf{a (f)}}, A part of static configuration with the square-rhombus (square-triangle) tiling. The radius of the disks shown here are $28\%$ of the actual value. Note that shadowed pentagons prevail and are essential to construct the tiling. {\bf{b}} and {\bf{c}} ({\bf{g}} and {\bf{h}}), Diffraction patterns and density profiles calculated from particles (top) and from pentagons and particles (bottom). {\bf{d (i)}}, Particle trajectories during a time interval of $10^5$ with the arrows pointing to the original direction of motion. {\bf{e} (j)}, van Hove autocorrelation function $G_a\left(\vec{r},t\right)$ at $t=6000$.
}
\end{figure*}

Figure~2a shows a part of static configuration of an OQC, which is rich of octagons and pentagons. The diffraction pattern shown in the top panel of Fig.~2b contains discrete sharp Bragg peaks with 8-fold symmetry, similar to that of the OQC of Cr-Ni-Si alloys \cite{kuo}. The density profile shown in the top panel of Fig.~2c further confirms the loss of density periodicity.

A close look at Fig.~2a reveals that each pentagon is surrounded by eight disks, which form a nice octagon. This implies that pentagons may be important structure elements in forming OQCs in our systems. Here we employ a polygonal order parameter $\delta={\rm max}\{\left|e_i/\bar{e}-1\right|\}$ ($i=1,2,...,5$) to numerically identify pentagons, where $e_i$ is the distance between the center of mass and vertex $i$ of a 5-sided polygon, and $\bar{e}=\sum_{i=1}^5 e_i/5$. Only 5-sided polygons with $\delta<0.1$ are identified as pentagons. By connecting centers of non-edge-adjacent pentagons, Fig.~2a shows that the OQC can be tessellated by 45$^\circ$ rhombi and squares. The number ratio of squares to rhombi is approximately $0.701$, close to $1:\sqrt{2}$ for perfect OQCs \cite{watanabe}. As shown in the bottom panel of Fig.~2b, by treating pentagons as units, the Bragg peaks become much sharper than those for single particles. Therefore, better quasicrystalline order is achieved by pentagons.

In addition to structures, the quasicrystalline order and significance of pentagons can be further verified from dynamics. Figure~2d shows the trajectories of two randomly chosen particles in the OQC. The trajectories are composed of a chain of pentagon loops. A particle moves along edges of a pentagon for a long time and suddenly escape from the pentagon and form a new pentagon with other particles, corresponding to a phason flip, whose presence causes liquid-like diffusion in QCs \cite{dzugutov1995}. The pentagon loops further emphasize the importance of pentagons in our QCs.

Figure~2e shows the van Hove autocorrelation function $G_{a}(\vec{r},t)$ for the OQC, which quantifies the probability distribution that a particle has been displaced by $\vec r$ at time $t$. In an intermediate time regime ($t=6000$ here), particles exhibit clear heterogeneous displacement. There are particles vibrating around their equilibrium positions, forming the central peak in $G_{a}(\vec{r},t)$ at $\vec{r}=0$. Surrounding the central peak are satellite peaks with 8-fold symmetry, consistent with the QC symmetry shown in structures.

Figures~2f-2j show the same structural and dynamical information for a DDQC. Interestingly, pentagons are still remarkable. As shown in Fig.~2f, each pentagon is surrounded by twelve disks sitting on vertexes of a dodecagon. Again, by connecting centers of non-edge-adjacent pentagons, the whole DDQC can be tiled by squares and triangles. The number ratio of triangles to squares is about 2.283, close to the ideal value of $4/\sqrt{3}$ for perfect DDQCs \cite{kawamura}. The significance of pentagons can also be told from their effects on sharpening Bragg peaks and the formation of pentagon loops in particle trajectories.

Now there comes a question why QCs can survive in certain regimes of the phase diagram. Owing to the complex structures of our QCs, it is difficult to directly calculate their free energies. We instead compare in Fig.~3 the $T=0$ potential energy of QCs with that of crystalline solids next to them. We take two typical values of $\alpha$ as examples: $\alpha=2.0$ and $2.5$, which correspond to widely studied harmonic and Hertzian repulsions. In the density regimes where we find QCs, the corresponding QCs have the lowest potential energy. Because the structures of QCs are more random than those of crystals, it is plausible to assume that the entropy of thermal QCs is higher as well. Thus, at $T>0$, QCs should have a lower free energy than crystals and are stable enough to survive.

\begin{figure}
\includegraphics[width=0.49\textwidth]{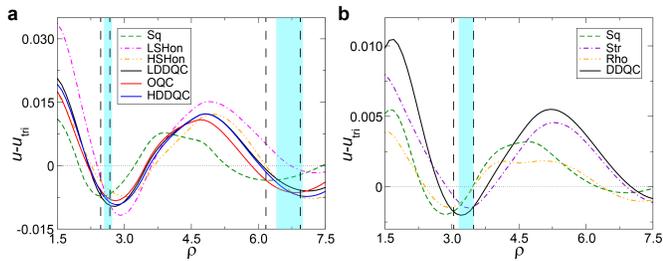}
\caption{\label{fig:fig3}  {\bf{Comparison of the potential energies of the $T=0$ QCs and neighboring solids}}. {\bf{a}} and $\bf{b}$ are for $\alpha=2.0$ and $\alpha=2.5$, respectively. The potential energy per particle $u$ is subtracted by that of a perfect triangular lattice, $u_{tri}$. The horizontal dotted line marks $u=u_{tri}$. The density regimes demarcated by the vertical dashed lines and light blue bands are where QCs have the lowest potential energy and where QCs exist, respectively. The absence of perfect match between the two may be due to the phase coexistence.
}
\end{figure}

All solid states shown in Fig.~1a are obtained by slowly quenching liquids below the melting temperature $T_m$. It has been proposed that prior to freezing some local orders may already start to develop in liquids \cite{tanaka}. Since pentagons are essential in our QCs, one may wonder whether a significant number of pentagons have already been formed in liquids. Moreover, the puzzling feature of our QCs is the lack of explicit competing length scales. It remains mysterious to us how the lengths established in Figs.~2a and 2f spontaneously emerge. To search for competing length scales in liquid states prior to the phase transition to QCs may provide us with some clues.

We thus compare structures of liquids at $T=1.1T_m$ over the whole range of densities of Fig.~1. The temperature envelop slightly above $T_m$ chosen here assures that the liquids stay at approximately the same distance away from the establishment of (quasi)crystalline order. In Fig.~4, we show the density dependence of the fraction of particles forming pentagons, $5N_{pentagon}/N$, and static structure factor, $S(k)$, for the liquids with harmonic and Hertzian repulsions, where $N_{pentagon}$ and $N$ denote the number of pentagons and total number of particles.

Figures.~4a and 4b indicate that pentagons have already accumulated in QC-forming liquids, leading to the maxima in $5N_{pentagon}/N$. The contour plots of $S(k)$ in Figs.~4c and 4d demonstrate two pronounced low-$k$ peaks in the density regimes where QCs reside. In Fig.~1a, there are two regimes of triangular solids. Beyond the maximum density concerned here, there are more regimes of triangular solids. The two peaks in $S(k)$ are apparently associated with the first peak of the liquids forming the two triangular solids on the lower and higher density sides of the QCs. It is the joint effects of high density and special capacity of the soft-core potentials to form multiple triangular solids that lead to the formation of pentagons, the building blocks of our QCs.

\begin{figure}
\includegraphics[width=0.49\textwidth]{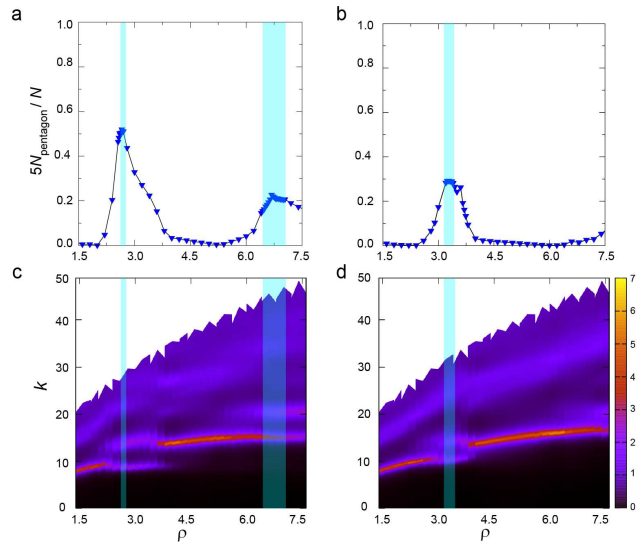}
\caption{\label{fig:fig4}  {\bf{Structural information in liquids}}. The left and right columns are for $\alpha =2.0$ and $2.5$, respectively. {\bf{a}} and {\bf{b}}, Density dependence of the fraction of disks forming pentagons, $5N_{pentagon}/N$. {\bf{c}} and {\bf{d}}, Contour plots of the static structure factors $S(k)$ with the color bar quantifies the value of $S(k)$. The liquids characterized here are slightly above the melting temperature at $T=1.1T_m$. The shadowed density regimes with light blue color are where QCs exist.
}
\end{figure}

The most surprising aspect of this work is the finding of a new class of soft QCs in so simple systems without any explicit multiple length scales. According to existing theories, QCs found here are unexpected. Thus, their existence poses a challenge to theories. Although we observe competing lengths in QC-forming liquids, the length scales are not those established in QCs. More in-depth studies are required to settle the underlying mechanisms of the spontaneous formation of the QC length scales. To track the microscopic pathways of the QC nucleation from supercooled liquids may be a necessary and direct approach toward this goal.

The soft-core potentials employed here have considerable theoretical merit \cite{liu}, which can also mimic particle interactions in experimental systems such as poly N-isopropylacrylamide colloids, granular materials, and foams \cite{zhang,majmudar,desmond}. High densities always bring us surprises with such potentials \cite{xu,miyazaki}. Now we even see QCs there. Implied by Fig.~1a, a long-range and relatively harder inter-particle repulsion (with smaller $\alpha$) needs to be modulated to verify our findings in experiments. Moreover, for both QCs found here, a pentagon surrounded by a $n$-side polygon forms the structural unit, which provides a promising motif to design $n$-fold QCs.

{\noindent \bf{Methods}}

Our systems are two-dimensional square boxes with side length $L$.  Periodic boundary conditions are applied in both directions. The system contains $N$ monodisperse disks with mass $m$. The units of energy, length, and mass are $\epsilon$, $\sigma$, and $m$. The time and temperature are in units of $\sqrt{m\sigma^2/\epsilon}$ and $\epsilon / k_B$ with $k_B$ being the Boltzmann constant. In this work, we mainly study $N=10000$ and $4096$ systems.

We perform molecular dynamics (MD) simulations in both the \emph{NVT} and \emph{NPT} ensembles. To outline the phase diagram, we slowly quench high-temperature liquids until solids are formed. We have verified that the quench rates are slow enough so that the phase boundaries are not sensitive to the change of the quench rate. To make sure that systems are in equilibrium, we first relax the system for a long time ($5\times 10^9$ MD steps with a time step $\Delta t = 0.01$ for solid states and $10^8$ MD steps for liquid states) and then collect data in the following $10^8$ MD steps. To get the static configurations shown in Figs.~1 and 2, we directly quench the equilibrium solid states to $T=0$ using the fast inertial relaxation engine algorithm \cite{bitzek}.

The diffraction patterns and density profiles are calculated from the static structure factor and radial distribution function, respectively:
$S(\vec{k})=\frac{1}{N}\left<\rho(\vec{k})\rho(-\vec{k})\right>$ and $g(\vec{r})=\frac{L^2}{2N^2}\left<\sum_{i=1}^{N}\sum_{j\neq{i}}^{N}\delta(\vec{r}-\vec{r}_{ij})\right>$, where $\rho(\vec{k})=\sum_{i=1}^{N}e^{i\vec{k}\cdot\vec{r}_{i}}$ is the Fourier transform of the density with $\vec{r}_i$ being the location of disk $i$, $\vec{k}$ is the wave vector satisfying the periodic boundary conditions, $\vec{r}_{ij}=\vec{r}_i -\vec{r}_j$ is the separation between disks $i$ and $j$, the sums are over all disks, and $\left< .\right>$ denotes the time average. The van Hove autocorrelation function is calculated from $G_a(\vec{r},t)=\left<\frac{1}{N}\sum_i \delta[\vec{r}-\vec{r}_i(t)+\vec{r}_i(0)]\right>$, where $\left<.\right>$ denotes the ensemble average and the sum is over all particles.

\end{document}